\documentclass[conference]{IEEEtran}
\IEEEoverridecommandlockouts
\usepackage{bm}
\usepackage{cite}
\usepackage{xcolor}
\usepackage{subfig}
\usepackage{caption}
\usepackage{makecell}
\usepackage{graphicx}
\usepackage{textcomp}
\usepackage{hyperref}
\usepackage{multirow}
\usepackage{enumerate}
\usepackage{algorithm}
\usepackage{tikz}
\usepackage{fancyhdr}

\usepackage{algpseudocode}
\usepackage{amsmath,amssymb,amsfonts}

\def\BibTeX{{\rm B\kern-.05em{\sc i\kern-.025em b}\kern-.08em
    T\kern-.1667em\lower.7ex\hbox{E}\kern-.125emX}}

\begin{document}

%~~~~~~~~~~~~~~~~~~~~~~~~~~~~~~~~~~~~~~~~~~~~~~~~~~~~~~~~~
\title{Boundary Value Test Input Generation Using Prompt Engineering with LLMs: Fault Detection and Coverage Analysis
}

\author{
\IEEEauthorblockN{
Xiujing Guo\IEEEauthorrefmark{1}, 
Chen Li\IEEEauthorrefmark{2}, 
Tatsuhiro Tsuchiya\IEEEauthorrefmark{1}}

\IEEEauthorblockA{\IEEEauthorrefmark{1}Graduate School of Information Science and Technology, Osaka University, Osaka, Japan\\
\{guoxiujing, t-tutiya\}@ist.osaka-u.ac.jp}
\IEEEauthorblockA{\IEEEauthorrefmark{2}Graduate School of Informatics, Nagoya University, Nagoya, Japan\\
li.chen.z2@a.mail.nagoya-u.ac.jp}
}

\maketitle

%~~~~~~~~~~~~~~~~~~~~~~~~~~~~~~~~~~~~~~~~~~~~~~~~~~~~~~~~~

%\newcommand\submittedtext{
  %\footnotesize This work has been submitted to the IEEE for possible publication. Copyright may be transferred without notice, after which this version may no longer be accessible.}

\begin{abstract}
\footnote{This work has been submitted to the IEEE for possible publication. Copyright may be transferred without notice, after which this version may no longer be accessible.}
As software systems grow more complex, automated testing has become essential to ensuring reliability and performance. Traditional methods for boundary value test input generation can be time-consuming and may struggle to address all potential error cases effectively, especially in systems with intricate or highly variable boundaries. This paper presents a framework for assessing the effectiveness of large language models (LLMs) in generating boundary value test inputs for white-box software testing by examining their potential through prompt engineering. Specifically, we evaluate the effectiveness of LLM-based test input generation by analyzing fault detection rates and test coverage, comparing these LLM-generated test sets with those produced using traditional boundary value analysis methods. Our analysis shows the strengths and limitations of LLMs in boundary value generation, particularly in detecting common boundary-related issues. However, they still face challenges in certain areas, especially when handling complex or less common test inputs. This research provides insights into the role of LLMs in boundary value testing, underscoring both their potential and areas for improvement in automated testing methods.
\end{abstract}

%~~~~~~~~~~~~~~~~~~~~~~~~~~~~~~~~~~~~~~~~~~~~~~~~~~~~~~~~~
\section{Introduction}
\label{sec:intro}
% Software testing
%As software systems continue to expand in scale and complexity, the challenge of ensuring reliability and optimal performance has become both critical and demanding \cite{chen2022software,dvivedi2024comparative}. Software testing, a vital aspect of development and quality assurance, plays a central role in identifying and resolving issues before they affect end users \cite{bertolino2007software}. Through systematic examination of software behavior across a variety of inputs and environments, testing seeks to validate functionality, security, and performance, thus minimizing the risk of production failures \cite{demir2024implementing}. The significance of effective testing is underscored by the increasing dependence on software in critical applications across various domains, including healthcare, finance, and transportation, where even minor flaws can lead to severe consequences. The cost of software failures is not only financial; it can also result in reputational damage and loss of user trust, emphasizing the need for rigorous testing practices \cite{hou2023systematic}. However, as systems grow more complex, traditional testing approaches struggle to efficiently and comprehensively cover large, intricate codebases \cite{rahman2024pioneering}. Exhaustive testing, which covers all possible input scenarios, is theoretically ideal but practically infeasible due to high time and resource costs \cite{guo2020automated}. This shows the need for efficient, flexible test input generation for effective software testing.

% Traditional methods
A foundational component of software testing is test input generation, which involves creating input values that reflect various conditions the software may encounter during deployment \cite{dustin1999automated}. Effective test inputs are essential for identifying edge cases, where software often exhibits unexpected behavior. Traditional input generation methods, such as Boundary Value Analysis (BVA), have been widely utilized to address this need \cite{puspitasari2023analysis}. BVA focuses on testing at the “boundaries” of input domains, which are known to harbor many potential errors \cite{forgacs2023domain}. By targeting these boundary values, BVA aims to identify off-by-one errors, overflow issues, and other edge case vulnerabilities. Despite its effectiveness in simpler contexts, BVA encounters limitations in managing the growing complexity of modern software systems, particularly when input spaces become less structured or when boundaries are more ambiguous or numerous \cite{ashiq2024challenges}. This limitation requires complementary approaches to cover more edge cases, ensuring thorough software validation.

% LLMs
The emergence of Large Language Models (LLMs) presents promising avenues for enhancing traditional software testing techniques. Models such as GPT \cite{floridi2020gpt} and BERT \cite{devlin2018bert}, trained on massive text corpora, have demonstrated strong capabilities in natural language understanding, text generation, and basic reasoning. These versatile models have inspired research into their applications beyond language tasks, including code generation \cite{svyatkovskiy2020intellicode}, test automation \cite{thummalapenta2012automating}, and broader software engineering practices \cite{borst2024protocol}. Recent studies have specifically investigated how LLMs can be harnessed to generate test cases for software applications, utilizing their capacity to understand syntax, semantics, and contextual relationships. By employing prompt engineering—crafting targeted queries or instructions—we can guide LLMs to produce inputs that align with boundary-focused testing strategies like BVA, enhancing traditional methods with new levels of adaptability \cite{santos2024we}.

% Motivations
Unlike BVA, which is primarily effective in well-defined numerical or categorical input domains, LLMs can dynamically generate diverse and contextually relevant inputs across a broader spectrum of scenarios \cite{wang2024software}. This capability is particularly valuable for exploring edge cases and identifying failure conditions that might otherwise go unnoticed with conventional methods alone. The flexibility and creativity of LLMs position them as powerful supplements to BVA, especially in cases where software systems present complex, multi-layered input structures or unconventional boundary conditions. Moreover, their ability to adaptively generate inputs based on contextual clues allows for a more nuanced exploration of potential vulnerabilities in software \cite{hou2023large}.

Motivated by these unique capabilities, this study investigates the potential of LLMs to generate boundary-focused test inputs that can effectively augment traditional techniques. We aim to understand how LLMs can contribute to error detection, enhance test coverage, and produce adaptable inputs that align with both conventional boundary values and unconventional edge cases. Specifically, our research evaluates LLM-generated inputs for their effectiveness in boundary-focused scenarios, measuring their contributions through metrics such as fault detection rates and test coverage. Through this investigation, we seek to illuminate both the strengths and limitations of LLMs in software testing, as well as their potential for integration with existing methodologies. The primary contributions of this study are as follows:
\begin{itemize} 
\item {\bf Generate boundary value test inputs using prompt-engineered interactions with LLMs:} We develop and implement techniques to create boundary value test inputs by leveraging LLMs.

\item {\bf Examination of fault detection rate and coverage:} We investigates the fault detection rates and coverage metrics (statement and branch coverage) for test inputs generated by LLMs, providing insights into the effectiveness of boundary value testing.

\item {\bf Comprehensive data analysis:} Various analyses were conducted on the generated test inputs, including comparisons with traditional methods, regression analysis of coverage and fault detection, and the impact of test input quantity, offering a thorough evaluation of LLM-generated test inputs for boundary value analysis.

\item {\bf Provide insights and recommendations for future work on improving automated boundary test input generation using LLMs:} We offer actionable insights and recommendations aimed at enhancing automated boundary test input generation processes with LLMs. 
\end{itemize}

%The paper is organized as follows: Section II reviews related work, Section III details the proposed approach, Section IV presents experiments and results, and Section V provides the conclusion and future work.
The remaining parts of the paper are organized as follows. Section \ref{sec:related} is dedicated to the review of related work. Section \ref{sec:method} describes the details of the proposed approach. In section \ref{sec:exp}, the experiments and results are presented. Section \ref{sec:conclusion} includes conclusion and some directions for future work.

%The remainder of this paper is structured as follows: Section \ref{sec:related} provides a comprehensive review of the existing literature on automated test input generation, focusing on the integration of LLMs within software testing practices. Section \ref{sec:method} outlines the methodology employed for generating test inputs, detailing the processes used to evaluate their effectiveness. Section \ref{sec:exp} presents the experimental results, including an analysis of error detection rates and branch coverage metrics. Section \ref{sec:discuss} discusses the implications of the findings, along with their significance and any limitations identified during the research. Finally, Section \ref{sec:conclusion} concludes the paper, summarizing the key contributions and proposing avenues for future research.

%~~~~~~~~~~~~~~~~~~~~~~~~~~~~~~~~~~~~~~~~~~~~~~~~~~~~~~~~~
\section{Related Work}
\label{sec:related}
%The field of software testing has witnessed considerable advancements in response to the complexities of modern software systems. This section reviews related work in three key areas: traditional test input generation methods, the integration of LLMs in software testing, and the current challenges and opportunities associated with these developments.

%~~~~~~~~~~~~~~~~~~~~~~~~~~~
\subsection{Traditional Test Input Generation Methods} 
Test input generation is a critical component of software testing, aimed at creating inputs that effectively validate the behavior of software applications. Among the traditional techniques, BVA has long been a standard approach. BVA focuses on testing at the boundaries of input domains, as these regions often contain potential error-prone areas, such as off-by-one errors and overflow issues \cite{puspitasari2023analysis,forgacs2023domain}. The primary advantage of BVA lies in its simplicity and effectiveness in revealing common types of errors related to boundary conditions. However, it struggles with complex systems where input domains are less structured or when boundaries are ambiguous \cite{dobslaw2020boundary}.

Equivalence partitioning is another well-known method, which divides input data into distinct classes that are expected to behave similarly \cite{reid1997empirical}. This method minimizes the number of test cases required while maximizing coverage across different input scenarios. While Equivalence partitioning is effective in reducing redundancy in test cases, it may overlook specific edge cases that could lead to failures. Random testing is also employed in software testing, where inputs are generated randomly within specified bounds \cite{arcuri2011random}. This method can cover a wide range of scenarios without the need for comprehensive domain knowledge. However, the randomness can lead to inconsistency in coverage, and it may require a substantial number of iterations to encounter rare edge cases. Additionally, the lack of structured approach often results in inefficient test cases that do not effectively target potential faults. Other approaches include model-based testing \cite{ferdous2023evombt}, which utilizes formal models to generate test cases based on the expected behavior of the system. This method provides thorough coverage of the software but often requires significant initial effort in creating and maintaining the models. The complexity involved makes it impractical for rapidly evolving software systems.

%~~~~~~~~~~~~~~~~~~~~~~~~~~~
\subsection{LLMs in Software Testing} 
The advent of LLMs such as GPT-3 \cite{floridi2020gpt} and BERT \cite{devlin2018bert} has significantly reshaped the field of software testing by introducing innovative methods for test input generation. These models, which have been trained on extensive corpora of text, exhibit remarkable capabilities in natural language processing \cite{naveed2023comprehensive}, question answering \cite{hu2023tifa}, and even drug discovery \cite{li2024tengan},\cite{chen2024gxvaes}. Researchers are exploring ways to use these capabilities to create effective test cases for software applications.

Studies have shown that LLMs can generate contextually relevant inputs by leveraging their understanding of syntax, semantics, and domain-specific knowledge \cite{ciatto2024large}. By employing prompt engineering techniques, developers can direct LLMs to create inputs that align with specific testing strategies. For example, recent studies demonstrate the ways in which LLMs can dynamically generate a wide range of test cases, covering not only typical inputs but also edge cases and unconventional scenarios often missed by traditional methods \cite{boukhlif2024towards,chen2024chatunitest}. This adaptability positions LLMs as powerful supplements to existing testing frameworks, offering not only enhanced coverage but also improved detection of failure conditions, especially in areas where conventional testing may fall short. By generating diverse test cases, LLMs help uncover hidden issues and ensure robust software performance.

One significant advantage of using LLMs is their ability to generate complex input structures that may not fit neatly into traditional categories \cite{yang2024harnessing}. This can be particularly useful in modern software applications that often handle multifaceted user interactions and diverse data types. Furthermore, LLMs can continuously improve their output quality as they are trained on larger datasets, which can lead to better understanding and generation of contextually appropriate inputs \cite{patil2024review}.

%~~~~~~~~~~~~~~~~~~~~~~~~~~~
\subsection{Challenges and Opportunities} 
Traditional BVA has generally focused on black-box testing, where boundaries are identified through techniques like equivalence partitioning. These methods typically rely on the availability of a specification to determine boundary information. However, in today’s open-source era, many projects lack formal software specifications, which complicates the process of identifying boundaries. This shift presents challenges for traditional BVA, as it must adapt to scenarios where boundary values need to be inferred directly from the code, rather than from well-defined specifications.
This paper considers the BVA in white-box testing. In white-box testing, boundary value analysis (BVA) seeks to uncover input values that provoke changes in the execution path, targeting conditions that can lead to distinct program states. However, identifying these boundary conditions in white-box testing presents unique challenges. Unlike black-box BVA, which assumes input variables are independent and derives boundaries from combinations of each variable’s limit, white-box BVA must account for inter-dependencies among variables within the code. For instance, conditions like \( a + b \leq 5 \) mean that boundaries cannot be isolated to a single variable; instead, they depend on satisfying complex relationships between variables. This complexity introduces a significant challenge in defining boundary values that are feasible under actual program constraints.

A primary difficulty in white-box BVA lies in identifying and generating test inputs that not only reach specific program paths but also effectively test boundaries. Traditional approaches to solving this problem, like using SMT solvers \cite{zhang2015boundary}, have been limited by scalability issues; SMT solvers struggle with large, complex codebases and multi-variable dependencies. LLMs, on the other hand, offer a unique potential to tackle these complex boundary generation tasks due to their ability to model and synthesize language-based patterns from extensive code data. However, the current focus of LLMs has been on achieving high coverage rather than identifying critical boundary values. This gap signifies an opportunity for LLMs to move beyond general input generation toward boundary-specific analysis.

To adapt LLMs for boundary value generation in white-box testing, several key opportunities exist. First, enhancing prompt engineering to encourage the model to generate inputs specifically at or near boundary conditions could refine LLMs’ BVA effectiveness. Second, using fine-tuning techniques with annotated boundary examples could improve LLMs' ability to generate inputs that precisely capture the intended boundary criteria. Additionally, integrating constraints handling within the LLM framework—potentially combining the LLM’s language processing strengths with lightweight constraint-solving methods—could address complex conditional boundaries while maintaining scalability.

LLMs’ adaptability to a wide range of programming languages and patterns also opens doors for them to support scalable, path-specific BVA, allowing for exploration of diverse boundary conditions across codebases. As LLMs continue to evolve, they hold significant promise for not only achieving coverage but also effectively identifying edge cases critical to robust software testing, marking an exciting frontier in white-box testing research.

%~~~~~~~~~~~~~~~~~~~~~~~~~~~~~~~~~~~~~~~~~~~~~~~~~~~~~~~~~
\section{Methodology}
\label{sec:method}

This section introduces a framework designed to evaluate the effectiveness of LLMs in generating boundary test inputs. The framework incorporates a series of prompt-based test generation strategies, where specific prompts guide the LLM to produce test inputs with boundary characteristics. By varying the prompt design to control the type of test inputs generated (e.g., boundary-focused or general inputs), we evaluate the quality of generated test inputs based on metrics like coverage and fault detection rates. By comparing the outcomes across different prompts, this framework enables a detailed assessment of the strengths and limitations of LLMs in boundary test input generation, providing insights for future prompt optimization.

%This section introduces our methodology in detail, covering the process of crafting prompts to assess LLMs, generating test sets based on these prompts, the approach to testing and evaluating LLM performance, and the metrics used for performance measurement.

%~~~~~~~~~~~~~~~~~~~~~~~~~~~
%\subsection{Overview of the Testing Flow} 
\subsection{Framework Overview for Evaluating Boundary Input Generation Using LLMs}
\begin{figure*}[t]
\centering
\includegraphics[width=0.75\hsize]{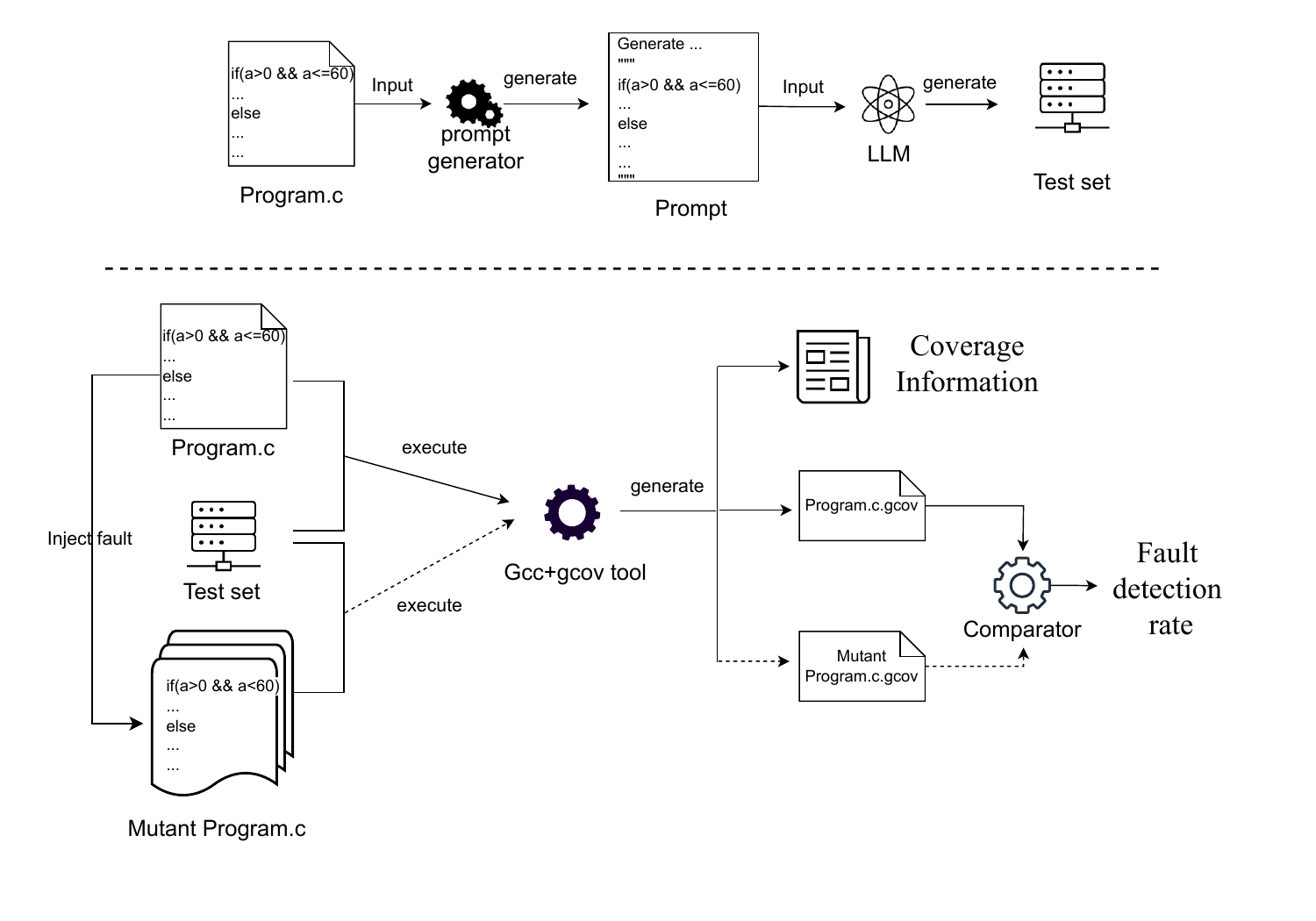}
\caption{Framework for evaluating LLM-based boundary test input generation using prompt engineering.}
\label{fig:fig2}
\end{figure*}

Figure \ref{fig:fig2} provides an overview of the architectural flow for generating boundary value test inputs through prompt engineering with LLMs. Our testing methodology we employ follows a structured flow, which begins with the preparation of relevant prompts for the LLMs and progresses through multiple stages to generate and evaluate test sets. The process is divided into the following key steps: prompt generation, test input generation using LLMs, mutation of the program under test, execution path extraction, and evaluation of fault detection and code coverage.
\begin{itemize}
\item[1)~] {\bf Prompt generation:} The process begins by creating prompts tailored to the specific program under test (e.g., C or C++ code). These prompts are used to instruct the LLMs on how to generate test inputs that cover a range of scenarios, including edge cases and boundary conditions.

\item[2)~] {\bf Test input generation:} With the prompts in hand, the LLMs, such as GPT-4o, generates a test set composed of multiple test inputs. These test inputs are designed to assess various program behaviors under different conditions, ensuring comprehensive test coverage and boundary coverage.

\item[3)~] {\bf Mutation of the program:} In this step, we manually introduce six types of faults into the original program to create mutant versions. This process simulates real-world software bugs and allows us to evaluate the effectiveness of the generated test inputs in detecting faults.

\item[4)~] {\bf Execution path extraction:} Using the original program, the generated test set, and the mutant programs, we execute the Gcov tool to extract detailed execution paths. This provides a clear picture of which paths in the code are covered by the test set and helps identify areas that may need additional testing.

\item[5)~] {\bf Evaluation:} Finally, we evaluate the performance of the generated test inputs using two main metrics: fault detection rate and coverage information. The fault detection rate quantifies how effectively the test inputs identify faults, while coverage information—extracted from the Gcov tool—provides insights into the extent of code exercised by the tests.
\end{itemize}

%~~~~~~~~~~~~~~~~~~~~~~~~~~~
\subsection{Prompt Engineering} 
Prompt engineering is a critical step in guiding the LLM to generate useful and accurate test inputs. The quality of the prompts directly impacts the effectiveness of the test inputs produced. In our methodology, we generate prompts that specify the type of input values required (e.g., boundary values), the expected behavior of the program, and the specific context in which the test inputs are to be applied.

The prompt needs to specify relevant information such as:
\begin{itemize}
\item The type of program being tested (C or C++).
\item The specific boundary conditions or edge cases that need to be considered.
\item Constraints or requirements for the input data (e.g., number).
\end{itemize}
As described in the next section, we created and tested simple prompts incorporating the above information at this stage of our study. Refining prompts using advanced prompt engineering techniques, such as RAG (Retrieval-Augmented Generation), is left for future work.

%For example, a typical prompt might be “generate test cases for a C program that implements a sorting algorithm. Focus on testing boundary values such as the smallest, largest, and middle values. Ensure that edge cases, such as an empty list or a list with identical elements, are also considered.” A prompt for boundary value testing can be represented as:
%\begin{align}
%\textit{Prompt} = ~&\textit{Generate test cases for a program that sorts } \\\nonumber
%&\textit{an array of integers},
%\end{align}
%where the prompt dynamically includes conditions to cover boundary values such as:
%\begin{align}
%\textit{Boundary}\in \{Min, Max, Mid\}
%\end{align}
%This ensures that the model understands the necessary scope and generates relevant test cases.

%~~~~~~~~~~~~~~~~~~~~~~~~~~~
\subsection{Fault Detection Ability}
In our experiment, we systematically modify specific statements in the source code to introduce faults, resulting in multiple faulty versions of the program, which we refer to as mutants. This strategy allows us to assess the effectiveness of the generated test inputs in detecting these deliberately introduced errors. 
For each generated set of test inputs, we execute the entire set against each faulty version of the program and record the number of killed mutants. For example, if we use an LLM to generate a test set consisting of 50 test inputs for a given program, we execute each faulty version of the program against these 50 inputs. If at least one of these inputs can expose an inserted fault, the test set is considered to have killed that mutant. 

%The kill rate is then calculated as follows:
%\begin{equation}
%kill\_rate=\frac{the\ number\ of\ killed\ mutants}{total\ number\ of\ mutants}. \nonumber
%\end{equation}

We introduce six fault types into the program under test through manual injection. Among these, off-by-one bugs (OBOBs) are a common boundary fault, occurring when a computation mistakenly uses a value that is one increment above or below the correct one \cite{zhang2015boundary}. The other fault types are based on five widely used mutation operators \cite{jia2010analysis}: constant replacement (CR), relational operator replacement (ROR), arithmetic operator replacement (AOR), scalar variable replacement (SVR), and logical operator replacement (LOR).

In black-box testing, determining whether a fault has been eliminated typically relies on comparing the program’s outputs across various test cases. Even after executing all test cases in a test set , some “surviving” mutants may remain. To improve the effectiveness of the test set, a tester can add further test inputs to target these remaining mutants. However, certain mutants, known as Equivalent Mutants, cannot be killed as they produce the same output as the original program, despite their syntactic differences. These mutants are functionally identical to the original program \cite{jia2010analysis}.

Considering this challenge, in white-box testing, we evaluate fault detection by comparing program execution path information. White-box boundary value testing focuses on covering boundary values specifically for comparison predicates. Here, each atomic Boolean expression in the path condition is treated as a predicate, which could be Boolean variables, comparison predicates ($>, >=, <, <=, =, \neq $), etc., and should not contain any Boolean operator (such as  $\land, \vee, \neg $, etc.)\ \cite{zhang2015boundary}. Each comparison predicate contains two branches. For example, the comparison predicate $(a > 0)$ results in two possible branches: one where the condition is true $(a > 0)$ and one where the condition is false $(a <= 0)$. In white-box testing, boundaries are closely related to branches in the code. These branches represent different paths the program can take, and testing these boundaries helps ensure that edge cases are properly handled.

In this study, we use the Gcov tool \cite{Gcov} to extract execution paths, defining an execution path as the specific combination of branch executions in response to a given input. Gcov is a source code coverage analysis and statement-by-statement profiling tool released with GCC, capable of generating detailed counts of the number of executions per statement in a program. It annotates source code for statement coverage and branch coverage analysis in C/C++ files. When the program runs, Gcov produces a {\tt program.c.gcov} file, which contains the number of executions for each branch. We extract execution path information directly from this file for comparison. A mutant is considered ``killed" if at least one test input results in an execution path different from that of the correct version.

%~~~~~~~~~~~~~~~~~~~~~~~~~~~
\subsection{Algorithm} 
The overall algorithm for generating and evaluating the test set follows the flow described above, with a focus on the key steps involved in creating test inputs, executing them, and evaluating their effectiveness. The algorithm is demonstrated in Algorithm \ref{alg:alg}.
\begin{algorithm}[t]
\caption{Test set generation and evaluation algorithm}
\label{alg:alg}
\begin{algorithmic}[1]
\State \textbf{Input:} 
\State $P \gets$ Program under test (C or C++ code)
\State $T \gets$ Prompt generated for LLM input 
\State \textbf{Test Input Generation:} 
\State $S \gets \text{LLM}(T)$ \Comment{Generate test set based on prompt $T$}
\State \textbf{Fault Injection:} 
\State $M \gets \{m_1, m_2, \dots, m_k\}$ \Comment{Inject faults into $P$ to create mutant programs $M$}
\State \textbf{Execution Path and Coverage Extraction:}
\State \textmd{Execute $P$ and all $m_i$ in $M$ with every input in $S$, recording the path for every execution and the code coverage $C$ for $P$.} 
\State \textbf{Fault Detection:}
\State $Kill\_rate \gets \frac{\text{the number of killed mutants}}{\text{total number of mutants}} \times 100$ \Comment{Determine killed mutants by comparing paths.}
%\State $C \gets \frac{\text{Covered statement (branch)}}{\text{Total statement (branch)}} \times 100$
\State \textbf{Output:} 
\State $Kill\_rate$ (Fault detection rate)
\State $C$ (Coverage rate)
\State \text{Insights on LLM-generated test input effectiveness}
\end{algorithmic}
\end{algorithm}
This process allows to systematically assess the quality and effectiveness of the test inputs generated by the LLM and ensure comprehensive software validation through a combination of fault detection and coverage evaluation.

%~~~~~~~~~~~~~~~~~~~~~~~~~~~~~~~~~~~~~~~~~~~~~~~~~~~~~~~~~
\section{Experiments}
\label{sec:exp}

\begin{table}[t]
\caption{Details of 8 subject programs}
\label{tab:table1}
\resizebox{\linewidth}{!}{
\centering 
\begin{tabular}{|c|c|c|cccccc|c|}
\hline
\multirow{2}{*}{Program} & \multirow{2}{*}{Dim} & \multirow{2}{*}{Size(LOC)} & \multicolumn{6}{c|}{Fault types}                                                                                                           & \multirow{2}{*}{Total Faults} \\ \cline{4-9}
                         &                      &                            & \multicolumn{1}{c|}{CR} & \multicolumn{1}{c|}{AOR} & \multicolumn{1}{c|}{LOR} & \multicolumn{1}{c|}{ROR} & \multicolumn{1}{c|}{SVR} & OBOB &                               \\ \hline
nextDate                 & 3                    & 90                         & \multicolumn{1}{c|}{}   & \multicolumn{1}{c|}{}    & \multicolumn{1}{c|}{8}   & \multicolumn{1}{c|}{7}   & \multicolumn{1}{c|}{}    & 16   & 31                            \\ \hline
bessj                    & 2                    & 133                        & \multicolumn{1}{c|}{1}  & \multicolumn{1}{c|}{1}   & \multicolumn{1}{c|}{}    & \multicolumn{1}{c|}{11}  & \multicolumn{1}{c|}{4}   & 10   & 27                            \\ \hline
plgndr                   & 3                    & 65                         & \multicolumn{1}{c|}{}   & \multicolumn{1}{c|}{2}   & \multicolumn{1}{c|}{2}   & \multicolumn{1}{c|}{6}   & \multicolumn{1}{c|}{3}   & 6    & 19                            \\ \hline
triType                  & 3                    & 41                         & \multicolumn{1}{c|}{}   & \multicolumn{1}{c|}{2}   & \multicolumn{1}{c|}{6}   & \multicolumn{1}{c|}{7}   & \multicolumn{1}{c|}{1}   & 2    & 18                            \\ \hline
findMiddle               & 3                    & 36                         & \multicolumn{1}{c|}{}   & \multicolumn{1}{c|}{}    & \multicolumn{1}{c|}{5}   & \multicolumn{1}{c|}{14}  & \multicolumn{1}{c|}{}    &      & 19                            \\ \hline
expint                   & 2                    & 109                        & \multicolumn{1}{c|}{2}  & \multicolumn{1}{c|}{5}   & \multicolumn{1}{c|}{10}  & \multicolumn{1}{c|}{2}   & \multicolumn{1}{c|}{3}   & 12   & 34                            \\ \hline
tcas                     & 12                   & 182                        & \multicolumn{1}{c|}{4}  & \multicolumn{1}{c|}{}    & \multicolumn{1}{c|}{6}   & \multicolumn{1}{c|}{2}   & \multicolumn{1}{c|}{5}   & 17   & 34                            \\ \hline
English                  & 2                    & 25                         & \multicolumn{1}{c|}{7}  & \multicolumn{1}{c|}{2}   & \multicolumn{1}{c|}{10}  & \multicolumn{1}{c|}{1}   & \multicolumn{1}{c|}{3}   & 14   & 25                            \\ \hline
\end{tabular}
}
\end{table}

\begin{table}[t]
\caption{Experimental programs}
\label{tab:table2}
\centering 
\begin{tabular}{|c|c|}
\hline
Prog Name  & Description                                   \\ \hline
triType \cite{Williams2005}    & The type of a triangle                        \\ \hline
nextDate \cite{Awedikian2009}   & Calculate the following date of the given day \\ \hline
findMiddle \cite{Ghani2009} & Find the middle number among three numbers    \\ \hline
bessj \cite{Teukolsky1992}      & Bessel function J of general integer order    \\ \hline
expint \cite{Teukolsky1992}    & Exponential integral                          \\ \hline
plgndr \cite{Teukolsky1992}     & Legendre polynomials                          \\ \hline
tcas \cite{Do2005}       & Aircraft collision avoidance system           \\ \hline
English       & English examination program           \\ \hline
\end{tabular}
\end{table}

\begin{table*}[t]
\caption{Comparison of kill rate with RT, Klee, MLBVA, and GPT (n = number of generated test inputs)}
\label{tab:table3}
\centering 
\resizebox{0.8\linewidth}{!}{
\begin{tabular}{|c|cccccccc|}
\hline
\multirow{2}{*}{Method} & \multicolumn{8}{c|}{Kill Rate}\\\cline{2-9} 
                        & \multicolumn{1}{c|}{triType}                                               & \multicolumn{1}{c|}{nextDate}                                              & \multicolumn{1}{c|}{findMiddle}                                            & \multicolumn{1}{c|}{bessj}                                                 & \multicolumn{1}{c|}{expint}                                                & \multicolumn{1}{c|}{plgndr}                                                   & \multicolumn{1}{c|}{tcas}                                                    & English                                               \\ \hline
BVAGPT1                 & \multicolumn{1}{c|}{\begin{tabular}[c]{@{}c@{}}0.76\\ (n=10)\end{tabular}} & \multicolumn{1}{c|}{\begin{tabular}[c]{@{}c@{}}0.83\\ (n=29)\end{tabular}} & \multicolumn{1}{c|}{\begin{tabular}[c]{@{}c@{}}1.00\\ (n=12)\end{tabular}} & \multicolumn{1}{c|}{\begin{tabular}[c]{@{}c@{}}0.59\\ (n=9)\end{tabular}}  & \multicolumn{1}{c|}{\begin{tabular}[c]{@{}c@{}}0.88\\ (n=10)\end{tabular}} & \multicolumn{1}{c|}{\begin{tabular}[c]{@{}c@{}}0.68\\ (n=12)\end{tabular}}    & \multicolumn{1}{c|}{\begin{tabular}[c]{@{}c@{}}0.12\\ (n=5)\end{tabular}}    & \begin{tabular}[c]{@{}c@{}}0.88\\ (n=49)\end{tabular} \\ \hline
BVAGPT2                 & \multicolumn{1}{c|}{\begin{tabular}[c]{@{}c@{}}0.71\\ (n=10)\end{tabular}} & \multicolumn{1}{c|}{\begin{tabular}[c]{@{}c@{}}0.83\\ (n=14)\end{tabular}} & \multicolumn{1}{c|}{\begin{tabular}[c]{@{}c@{}}0.7\\ (n=10)\end{tabular}}  & \multicolumn{1}{c|}{\begin{tabular}[c]{@{}c@{}}0.48\\ (n=5)\end{tabular}}  & \multicolumn{1}{c|}{\begin{tabular}[c]{@{}c@{}}0.82\\ (n=15)\end{tabular}} & \multicolumn{1}{c|}{\begin{tabular}[c]{@{}c@{}}0.68\\ (n=21)\end{tabular}}    & \multicolumn{1}{c|}{\begin{tabular}[c]{@{}c@{}}0.12\\ (n=3)\end{tabular}}    & \begin{tabular}[c]{@{}c@{}}0.88\\ (n=10)\end{tabular} \\ \hline
BVAGPT3                 & \multicolumn{1}{c|}{\begin{tabular}[c]{@{}c@{}}0.61\\ (n=11)\end{tabular}} & \multicolumn{1}{c|}{\begin{tabular}[c]{@{}c@{}}0.96\\ (n=28)\end{tabular}} & \multicolumn{1}{c|}{\begin{tabular}[c]{@{}c@{}}0.82\\ (n=13)\end{tabular}} & \multicolumn{1}{c|}{\begin{tabular}[c]{@{}c@{}}0.62\\ (n=8)\end{tabular}}  & \multicolumn{1}{c|}{\begin{tabular}[c]{@{}c@{}}1.00\\ (n=17)\end{tabular}} & \multicolumn{1}{c|}{\begin{tabular}[c]{@{}c@{}}0.68\\ (n=12)\end{tabular}}    & \multicolumn{1}{c|}{\begin{tabular}[c]{@{}c@{}}0.25\\ (n=8)\end{tabular}}    & \begin{tabular}[c]{@{}c@{}}0.92\\ (n=14)\end{tabular} \\ \hline
genGPT1                    & \multicolumn{1}{c|}{\begin{tabular}[c]{@{}c@{}}0.61\\ (n=10)\end{tabular}} & \multicolumn{1}{c|}{\begin{tabular}[c]{@{}c@{}}0.83\\ (n=15)\end{tabular}} & \multicolumn{1}{c|}{\begin{tabular}[c]{@{}c@{}}0.85\\ (n=12)\end{tabular}} & \multicolumn{1}{c|}{\begin{tabular}[c]{@{}c@{}}0.48\\ (n=10)\end{tabular}} & \multicolumn{1}{c|}{\begin{tabular}[c]{@{}c@{}}0.90\\ (n=11)\end{tabular}} & \multicolumn{1}{c|}{\begin{tabular}[c]{@{}c@{}}0.73\\ (n=12)\end{tabular}}    & \multicolumn{1}{c|}{\begin{tabular}[c]{@{}c@{}}0.28\\ (n=10)\end{tabular}}   & \begin{tabular}[c]{@{}c@{}}0.71\\ (n=10)\end{tabular} \\ \hline
genGPT2                    & \multicolumn{1}{c|}{\begin{tabular}[c]{@{}c@{}}0.66\\ (n=10)\end{tabular}} & \multicolumn{1}{c|}{\begin{tabular}[c]{@{}c@{}}0.74\\ (n=10)\end{tabular}} & \multicolumn{1}{c|}{\begin{tabular}[c]{@{}c@{}}0.94\\ (n=10)\end{tabular}} & \multicolumn{1}{c|}{\begin{tabular}[c]{@{}c@{}}0.62\\ (n=12)\end{tabular}} & \multicolumn{1}{c|}{\begin{tabular}[c]{@{}c@{}}0.94\\ (n=13)\end{tabular}} & \multicolumn{1}{c|}{\begin{tabular}[c]{@{}c@{}}0.73\\ (n=13)\end{tabular}}    & \multicolumn{1}{c|}{\begin{tabular}[c]{@{}c@{}}0.25\\ (n=10)\end{tabular}}   & \begin{tabular}[c]{@{}c@{}}0.64\\ (n=10)\end{tabular} \\ \hline
genGPT3                    & \multicolumn{1}{c|}{\begin{tabular}[c]{@{}c@{}}0.61\\ (n=10)\end{tabular}} & \multicolumn{1}{c|}{\begin{tabular}[c]{@{}c@{}}0.74\\ (n=10)\end{tabular}} & \multicolumn{1}{c|}{\begin{tabular}[c]{@{}c@{}}0.97\\ (n=13)\end{tabular}} & \multicolumn{1}{c|}{\begin{tabular}[c]{@{}c@{}}0.62\\ (n=19)\end{tabular}} & \multicolumn{1}{c|}{\begin{tabular}[c]{@{}c@{}}0.88\\ (n=18)\end{tabular}} & \multicolumn{1}{c|}{\begin{tabular}[c]{@{}c@{}}0.73\\ (n=19)\end{tabular}}    & \multicolumn{1}{c|}{\begin{tabular}[c]{@{}c@{}}0.25\\ (n=5)\end{tabular}}    & \begin{tabular}[c]{@{}c@{}}0.68\\ (n=12)\end{tabular} \\ \hline
50BVAGPT1               & \multicolumn{1}{c|}{0.95}                                                  & \multicolumn{1}{c|}{0.7}                                                   & \multicolumn{1}{c|}{0.88}                                                  & \multicolumn{1}{c|}{0.59}                                                  & \multicolumn{1}{c|}{1.00}                                                  & \multicolumn{1}{c|}{0.73}                                                     & \multicolumn{1}{c|}{0.43}                                                    & 0.83                                                  \\ \hline
50BVAGPT2               & \multicolumn{1}{c|}{0.85}                                                  & \multicolumn{1}{c|}{0.96}                                                  & \multicolumn{1}{c|}{0.88}                                                  & \multicolumn{1}{c|}{0.51}                                                  & \multicolumn{1}{c|}{1.00}                                                  & \multicolumn{1}{c|}{0.73}                                                     & \multicolumn{1}{c|}{0.18}                                                    & 0.95                                                  \\ \hline
50BVAGPT3               & \multicolumn{1}{c|}{0.85}                                                  & \multicolumn{1}{c|}{0.87}                                                  & \multicolumn{1}{c|}{1.00}                                                  & \multicolumn{1}{c|}{0.70}                                                  & \multicolumn{1}{c|}{1.00}                                                  & \multicolumn{1}{c|}{0.73}                                                     & \multicolumn{1}{c|}{0.40}                                                    & 0.95                                                  \\ \hline
50genGPT1                  & \multicolumn{1}{c|}{0.80}                                                  & \multicolumn{1}{c|}{0.93}                                                  & \multicolumn{1}{c|}{1.00}                                                  & \multicolumn{1}{c|}{0.85}                                                  & \multicolumn{1}{c|}{0.88}                                                  & \multicolumn{1}{c|}{0.73}                                                     & \multicolumn{1}{c|}{0.46}                                                    & 0.92                                                  \\ \hline
50genGPT2                  & \multicolumn{1}{c|}{0.76}                                                  & \multicolumn{1}{c|}{0.87}                                                  & \multicolumn{1}{c|}{1.00}                                                  & \multicolumn{1}{c|}{0.77}                                                  & \multicolumn{1}{c|}{0.88}                                                  & \multicolumn{1}{c|}{0.73}                                                     & \multicolumn{1}{c|}{0.56}                                                    & 0.92                                                  \\ \hline
50genGPT3                  & \multicolumn{1}{c|}{0.90}                                                  & \multicolumn{1}{c|}{0.90}                                                  & \multicolumn{1}{c|}{1.00}                                                  & \multicolumn{1}{c|}{0.77}                                                  & \multicolumn{1}{c|}{0.94}                                                  & \multicolumn{1}{c|}{0.73}                                                     & \multicolumn{1}{c|}{0.34}                                                    & 0.95                                                  \\ \hline
RT(n=50)                & \multicolumn{1}{c|}{0.61}                                                  & \multicolumn{1}{c|}{0.58}                                                  & \multicolumn{1}{c|}{0.36}                                                  & \multicolumn{1}{c|}{0.52}                                                  & \multicolumn{1}{c|}{0.47}                                                  & \multicolumn{1}{c|}{0.63}                                                     & \multicolumn{1}{c|}{0.29}                                                    & 0.40                                                  \\ \hline
MLBVA(n=50)             & \multicolumn{1}{c|}{0.72}                                                  & \multicolumn{1}{c|}{0.70}                                                  & \multicolumn{1}{c|}{0.79}                                                  & \multicolumn{1}{c|}{0.85}                                                  & \multicolumn{1}{c|}{0.79}                                                  & \multicolumn{1}{c|}{0.95}                                                     & \multicolumn{1}{c|}{0.05}                                                    & 0.64                                                  \\ \hline
Klee                    & \multicolumn{1}{c|}{\begin{tabular}[c]{@{}c@{}}1.00\\ (n=14)\end{tabular}} & \multicolumn{1}{c|}{\begin{tabular}[c]{@{}c@{}}0.90\\ (n=56)\end{tabular}} & \multicolumn{1}{c|}{\begin{tabular}[c]{@{}c@{}}1.00\\ (n=13)\end{tabular}} & \multicolumn{1}{c|}{\begin{tabular}[c]{@{}c@{}}0.37\\ (n=2)\end{tabular}}  & \multicolumn{1}{c|}{\begin{tabular}[c]{@{}c@{}}0.38\\ (n=4)\end{tabular}}  & \multicolumn{1}{c|}{\begin{tabular}[c]{@{}c@{}}0.84\\ (n=14221)\end{tabular}} & \multicolumn{1}{c|}{\begin{tabular}[c]{@{}c@{}}1.00\\ (n=1290)\end{tabular}} & \begin{tabular}[c]{@{}c@{}}0.60\\ (n=8)\end{tabular}  \\ \hline
\end{tabular}
}
%\end{table*}
\vspace{10pt}
%\begin{table*}[htbp]
\caption{Comparison of statement coverage with RT, Concolic testing (Klee), MLBVA, and GPT.}
\label{tab:table4}
\centering 
\resizebox{0.8\linewidth}{!}{
\begin{tabular}{|c|cccccccc|}
\hline
\multirow{2}{*}{Method} & \multicolumn{8}{c|}{Statement Coverage}\\ \cline{2-9} 
                        & \multicolumn{1}{c|}{triType}                                               & \multicolumn{1}{c|}{nextDate}                                              & \multicolumn{1}{c|}{findMiddle}                                            & \multicolumn{1}{c|}{bessj}                                                 & \multicolumn{1}{c|}{expint}                                                & \multicolumn{1}{c|}{plgndr}                                                   & \multicolumn{1}{c|}{tcas}                                                    & English                                               \\ \hline
BVAGPT1                 & \multicolumn{1}{c|}{\begin{tabular}[c]{@{}c@{}}1.00\\ (n=10)\end{tabular}} & \multicolumn{1}{c|}{\begin{tabular}[c]{@{}c@{}}0.96\\ (n=29)\end{tabular}} & \multicolumn{1}{c|}{\begin{tabular}[c]{@{}c@{}}1.00\\ (n=12)\end{tabular}} & \multicolumn{1}{c|}{\begin{tabular}[c]{@{}c@{}}0.90\\ (n=9)\end{tabular}}  & \multicolumn{1}{c|}{\begin{tabular}[c]{@{}c@{}}0.87\\ (n=10)\end{tabular}} & \multicolumn{1}{c|}{\begin{tabular}[c]{@{}c@{}}0.89\\ (n=12)\end{tabular}}    & \multicolumn{1}{c|}{\begin{tabular}[c]{@{}c@{}}0.58\\ (n=5)\end{tabular}}    & \begin{tabular}[c]{@{}c@{}}1.00\\ (n=49)\end{tabular} \\ \hline
BVAGPT2                 & \multicolumn{1}{c|}{\begin{tabular}[c]{@{}c@{}}1.00\\ (n=10)\end{tabular}} & \multicolumn{1}{c|}{\begin{tabular}[c]{@{}c@{}}0.96\\ (n=14)\end{tabular}} & \multicolumn{1}{c|}{\begin{tabular}[c]{@{}c@{}}0.94\\ (n=10)\end{tabular}} & \multicolumn{1}{c|}{\begin{tabular}[c]{@{}c@{}}0.89\\ (n=5)\end{tabular}}  & \multicolumn{1}{c|}{\begin{tabular}[c]{@{}c@{}}0.75\\ (n=15)\end{tabular}} & \multicolumn{1}{c|}{\begin{tabular}[c]{@{}c@{}}0.89\\ (n=21)\end{tabular}}    & \multicolumn{1}{c|}{\begin{tabular}[c]{@{}c@{}}0.58\\ (n=3)\end{tabular}}    & \begin{tabular}[c]{@{}c@{}}1.00\\ (n=10)\end{tabular} \\ \hline
BVAGPT3                 & \multicolumn{1}{c|}{\begin{tabular}[c]{@{}c@{}}0.90\\ (n=11)\end{tabular}} & \multicolumn{1}{c|}{\begin{tabular}[c]{@{}c@{}}0.98\\ (n=28)\end{tabular}} & \multicolumn{1}{c|}{\begin{tabular}[c]{@{}c@{}}1.00\\ (n=13)\end{tabular}} & \multicolumn{1}{c|}{\begin{tabular}[c]{@{}c@{}}0.90\\ (n=8)\end{tabular}}  & \multicolumn{1}{c|}{\begin{tabular}[c]{@{}c@{}}0.87\\ (n=17)\end{tabular}} & \multicolumn{1}{c|}{\begin{tabular}[c]{@{}c@{}}0.89\\ (n=12)\end{tabular}}    & \multicolumn{1}{c|}{\begin{tabular}[c]{@{}c@{}}0.58\\ (n=8)\end{tabular}}    & \begin{tabular}[c]{@{}c@{}}1.00\\ (n=14)\end{tabular} \\ \hline
genGPT1                    & \multicolumn{1}{c|}{\begin{tabular}[c]{@{}c@{}}1.00\\ (n=10)\end{tabular}} & \multicolumn{1}{c|}{\begin{tabular}[c]{@{}c@{}}0.97\\ (n=15)\end{tabular}} & \multicolumn{1}{c|}{\begin{tabular}[c]{@{}c@{}}0.97\\ (n=12)\end{tabular}} & \multicolumn{1}{c|}{\begin{tabular}[c]{@{}c@{}}0.51\\ (n=10)\end{tabular}} & \multicolumn{1}{c|}{\begin{tabular}[c]{@{}c@{}}0.87\\ (n=11)\end{tabular}} & \multicolumn{1}{c|}{\begin{tabular}[c]{@{}c@{}}1.00\\ (n=12)\end{tabular}}    & \multicolumn{1}{c|}{\begin{tabular}[c]{@{}c@{}}0.92\\ (n=10)\end{tabular}}   & \begin{tabular}[c]{@{}c@{}}1.00\\ (n=10)\end{tabular} \\ \hline
genGPT2                    & \multicolumn{1}{c|}{\begin{tabular}[c]{@{}c@{}}1.00\\ (n=10)\end{tabular}} & \multicolumn{1}{c|}{\begin{tabular}[c]{@{}c@{}}0.92\\ (n=10)\end{tabular}} & \multicolumn{1}{c|}{\begin{tabular}[c]{@{}c@{}}1.00\\ (n=10)\end{tabular}} & \multicolumn{1}{c|}{\begin{tabular}[c]{@{}c@{}}0.91\\ (n=12)\end{tabular}} & \multicolumn{1}{c|}{\begin{tabular}[c]{@{}c@{}}0.87\\ (n=13)\end{tabular}} & \multicolumn{1}{c|}{\begin{tabular}[c]{@{}c@{}}1.00\\ (n=19)\end{tabular}}    & \multicolumn{1}{c|}{\begin{tabular}[c]{@{}c@{}}0.87\\ (n=5)\end{tabular}}    & \begin{tabular}[c]{@{}c@{}}1.00\\ (n=10)\end{tabular} \\ \hline
genGPT3                    & \multicolumn{1}{c|}{\begin{tabular}[c]{@{}c@{}}1.00\\ (n=10)\end{tabular}} & \multicolumn{1}{c|}{\begin{tabular}[c]{@{}c@{}}0.92\\ (n=10)\end{tabular}} & \multicolumn{1}{c|}{\begin{tabular}[c]{@{}c@{}}1.00\\ (n=13)\end{tabular}} & \multicolumn{1}{c|}{\begin{tabular}[c]{@{}c@{}}0.81\\ (n=19)\end{tabular}} & \multicolumn{1}{c|}{\begin{tabular}[c]{@{}c@{}}0.87\\ (n=18)\end{tabular}} & \multicolumn{1}{c|}{\begin{tabular}[c]{@{}c@{}}1.00\\ (n=19)\end{tabular}}    & \multicolumn{1}{c|}{\begin{tabular}[c]{@{}c@{}}0.92\\ (n=5)\end{tabular}}    & \begin{tabular}[c]{@{}c@{}}1.00\\ (n=12)\end{tabular} \\ \hline
50BVAGPT1               & \multicolumn{1}{c|}{1.00}                                                  & \multicolumn{1}{c|}{0.9}                                                   & \multicolumn{1}{c|}{0.97}                                                  & \multicolumn{1}{c|}{0.81}                                                  & \multicolumn{1}{c|}{0.87}                                                  & \multicolumn{1}{c|}{1.00}                                                     & \multicolumn{1}{c|}{0.92}                                                    & 1.00                                                  \\ \hline
50BVAGPT2               & \multicolumn{1}{c|}{1.00}                                                  & \multicolumn{1}{c|}{0.98}                                                  & \multicolumn{1}{c|}{0.97}                                                  & \multicolumn{1}{c|}{0.81}                                                  & \multicolumn{1}{c|}{0.87}                                                  & \multicolumn{1}{c|}{1.00}                                                     & \multicolumn{1}{c|}{0.58}                                                    & 1.00                                                  \\ \hline
50BVAGPT3               & \multicolumn{1}{c|}{1.00}                                                  & \multicolumn{1}{c|}{0.96}                                                  & \multicolumn{1}{c|}{1.00}                                                  & \multicolumn{1}{c|}{0.91}                                                  & \multicolumn{1}{c|}{0.87}                                                  & \multicolumn{1}{c|}{1.00}                                                     & \multicolumn{1}{c|}{0.93}                                                    & 1.00                                                  \\ \hline
50genGPT1                  & \multicolumn{1}{c|}{1.00}                                                  & \multicolumn{1}{c|}{0.97}                                                  & \multicolumn{1}{c|}{1.00}                                                  & \multicolumn{1}{c|}{0.91}                                                  & \multicolumn{1}{c|}{0.87}                                                  & \multicolumn{1}{c|}{1.00}                                                     & \multicolumn{1}{c|}{0.98}                                                    & 1.00                                                  \\ \hline
50genGPT2                  & \multicolumn{1}{c|}{1.00}                                                  & \multicolumn{1}{c|}{0.98}                                                  & \multicolumn{1}{c|}{1.00}                                                  & \multicolumn{1}{c|}{0.81}                                                  & \multicolumn{1}{c|}{0.87}                                                  & \multicolumn{1}{c|}{1.00}                                                     & \multicolumn{1}{c|}{0.98}                                                    & 1.00                                                  \\ \hline
50genGPT3                  & \multicolumn{1}{c|}{1.00}                                                  & \multicolumn{1}{c|}{0.98}                                                  & \multicolumn{1}{c|}{1.00}                                                  & \multicolumn{1}{c|}{0.91}                                                  & \multicolumn{1}{c|}{0.87}                                                  & \multicolumn{1}{c|}{1.00}                                                     & \multicolumn{1}{c|}{0.97}                                                    & 1.00                                                  \\ \hline
RT(n=50)                & \multicolumn{1}{c|}{0.9}                                                   & \multicolumn{1}{c|}{0.83}                                                  & \multicolumn{1}{c|}{0.86}                                                  & \multicolumn{1}{c|}{0.8}                                                   & \multicolumn{1}{c|}{0.72}                                                  & \multicolumn{1}{c|}{0.96}                                                     & \multicolumn{1}{c|}{0.97}                                                    & 0.96                                                  \\ \hline
MLBVA(n=50)             & \multicolumn{1}{c|}{0.95}                                                  & \multicolumn{1}{c|}{0.87}                                                  & \multicolumn{1}{c|}{0.97}                                                  & \multicolumn{1}{c|}{0.99}                                                  & \multicolumn{1}{c|}{0.86}                                                  & \multicolumn{1}{c|}{0.87}                                                     & \multicolumn{1}{c|}{0.58}                                                    & 0.96                                                  \\ \hline
Klee                    & \multicolumn{1}{c|}{\begin{tabular}[c]{@{}c@{}}1.00\\ (n=14)\end{tabular}} & \multicolumn{1}{c|}{\begin{tabular}[c]{@{}c@{}}1.00\\ (n=56)\end{tabular}} & \multicolumn{1}{c|}{\begin{tabular}[c]{@{}c@{}}1.00\\ (n=13)\end{tabular}} & \multicolumn{1}{c|}{\begin{tabular}[c]{@{}c@{}}0.37\\ (n=2)\end{tabular}}  & \multicolumn{1}{c|}{\begin{tabular}[c]{@{}c@{}}0.51\\ (n=4)\end{tabular}}  & \multicolumn{1}{c|}{\begin{tabular}[c]{@{}c@{}}1.00\\ (n=14221)\end{tabular}} & \multicolumn{1}{c|}{\begin{tabular}[c]{@{}c@{}}0.98\\ (n=1290)\end{tabular}} & \begin{tabular}[c]{@{}c@{}}1.00\\ (n=8)\end{tabular}  \\ \hline
\end{tabular}
}
\end{table*}

\begin{table*}[htbp]
\caption{Comparison of branch coverage with RT, Concolic testing (Klee), MLBVA, and GPT.}
\label{tab:table5}
\centering
\resizebox{0.8\linewidth}{!}{
\begin{tabular}{|c|cccccccc|}
\hline
\multirow{2}{*}{Method} & \multicolumn{8}{c|}{Branch Coverage}\\ \cline{2-9} 
                        & \multicolumn{1}{c|}{triType}                                               & \multicolumn{1}{c|}{nextDate}                                              & \multicolumn{1}{c|}{findMiddle}                                            & \multicolumn{1}{c|}{bessj}                                                 & \multicolumn{1}{c|}{expint}                                                & \multicolumn{1}{c|}{plgndr}                                                   & \multicolumn{1}{c|}{tcas}                                                    & English                                               \\ \hline
BVAGPT1                 & \multicolumn{1}{c|}{\begin{tabular}[c]{@{}c@{}}0.82\\ (n=10)\end{tabular}} & \multicolumn{1}{c|}{\begin{tabular}[c]{@{}c@{}}0.60\\ (n=29)\end{tabular}} & \multicolumn{1}{c|}{\begin{tabular}[c]{@{}c@{}}1.00\\ (n=12)\end{tabular}} & \multicolumn{1}{c|}{\begin{tabular}[c]{@{}c@{}}0.87\\ (n=9)\end{tabular}}  & \multicolumn{1}{c|}{\begin{tabular}[c]{@{}c@{}}0.94\\ (n=10)\end{tabular}} & \multicolumn{1}{c|}{\begin{tabular}[c]{@{}c@{}}0.86\\ (n=12)\end{tabular}}    & \multicolumn{1}{c|}{\begin{tabular}[c]{@{}c@{}}0.22\\ (n=5)\end{tabular}}    & \begin{tabular}[c]{@{}c@{}}1.00\\ (n=49)\end{tabular} \\ \hline
BVAGPT2                 & \multicolumn{1}{c|}{\begin{tabular}[c]{@{}c@{}}0.76\\ (n=10)\end{tabular}} & \multicolumn{1}{c|}{\begin{tabular}[c]{@{}c@{}}0.58\\ (n=14)\end{tabular}} & \multicolumn{1}{c|}{\begin{tabular}[c]{@{}c@{}}0.85\\ (n=10)\end{tabular}} & \multicolumn{1}{c|}{\begin{tabular}[c]{@{}c@{}}0.75\\ (n=5)\end{tabular}}  & \multicolumn{1}{c|}{\begin{tabular}[c]{@{}c@{}}0.72\\ (n=15)\end{tabular}} & \multicolumn{1}{c|}{\begin{tabular}[c]{@{}c@{}}0.86\\ (n=21)\end{tabular}}    & \multicolumn{1}{c|}{\begin{tabular}[c]{@{}c@{}}0.22\\ (n=3)\end{tabular}}    & \begin{tabular}[c]{@{}c@{}}0.95\\ (n=10)\end{tabular} \\ \hline
BVAGPT3                 & \multicolumn{1}{c|}{\begin{tabular}[c]{@{}c@{}}0.67\\ (n=11)\end{tabular}} & \multicolumn{1}{c|}{\begin{tabular}[c]{@{}c@{}}0.70\\ (n=28)\end{tabular}} & \multicolumn{1}{c|}{\begin{tabular}[c]{@{}c@{}}0.94\\ (n=13)\end{tabular}} & \multicolumn{1}{c|}{\begin{tabular}[c]{@{}c@{}}0.81\\ (n=8)\end{tabular}}  & \multicolumn{1}{c|}{\begin{tabular}[c]{@{}c@{}}0.94\\ (n=17)\end{tabular}} & \multicolumn{1}{c|}{\begin{tabular}[c]{@{}c@{}}0.86\\ (n=12)\end{tabular}}    & \multicolumn{1}{c|}{\begin{tabular}[c]{@{}c@{}}0.29\\ (n=8)\end{tabular}}    & \begin{tabular}[c]{@{}c@{}}1.00\\ (n=14)\end{tabular} \\ \hline
genGPT1                    & \multicolumn{1}{c|}{\begin{tabular}[c]{@{}c@{}}0.79\\ (n=10)\end{tabular}} & \multicolumn{1}{c|}{\begin{tabular}[c]{@{}c@{}}0.56\\ (n=15)\end{tabular}} & \multicolumn{1}{c|}{\begin{tabular}[c]{@{}c@{}}0.94\\ (n=12)\end{tabular}} & \multicolumn{1}{c|}{\begin{tabular}[c]{@{}c@{}}0.65\\ (n=10)\end{tabular}} & \multicolumn{1}{c|}{\begin{tabular}[c]{@{}c@{}}0.88\\ (n=11)\end{tabular}} & \multicolumn{1}{c|}{\begin{tabular}[c]{@{}c@{}}0.9\\ (n=12)\end{tabular}}     & \multicolumn{1}{c|}{\begin{tabular}[c]{@{}c@{}}0.64\\ (n=10)\end{tabular}}   & \begin{tabular}[c]{@{}c@{}}0.9\\ (n=10)\end{tabular}  \\ \hline
genGPT2                    & \multicolumn{1}{c|}{\begin{tabular}[c]{@{}c@{}}0.76\\ (n=10)\end{tabular}} & \multicolumn{1}{c|}{\begin{tabular}[c]{@{}c@{}}0.52\\ (n=10)\end{tabular}} & \multicolumn{1}{c|}{\begin{tabular}[c]{@{}c@{}}0.97\\ (n=10)\end{tabular}} & \multicolumn{1}{c|}{\begin{tabular}[c]{@{}c@{}}0.93\\ (n=12)\end{tabular}} & \multicolumn{1}{c|}{\begin{tabular}[c]{@{}c@{}}0.91\\ (n=13)\end{tabular}} & \multicolumn{1}{c|}{\begin{tabular}[c]{@{}c@{}}1.00\\ (n=13)\end{tabular}}    & \multicolumn{1}{c|}{\begin{tabular}[c]{@{}c@{}}0.57\\ (n=10)\end{tabular}}   & \begin{tabular}[c]{@{}c@{}}0.95\\ (n=10)\end{tabular} \\ \hline
genGPT3                    & \multicolumn{1}{c|}{\begin{tabular}[c]{@{}c@{}}0.76\\ (n=10)\end{tabular}} & \multicolumn{1}{c|}{\begin{tabular}[c]{@{}c@{}}0.52\\ (n=10)\end{tabular}} & \multicolumn{1}{c|}{\begin{tabular}[c]{@{}c@{}}0.97\\ (n=13)\end{tabular}} & \multicolumn{1}{c|}{\begin{tabular}[c]{@{}c@{}}0.87\\ (n=19)\end{tabular}} & \multicolumn{1}{c|}{\begin{tabular}[c]{@{}c@{}}0.91\\ (n=18)\end{tabular}} & \multicolumn{1}{c|}{\begin{tabular}[c]{@{}c@{}}0.9\\ (n=19)\end{tabular}}     & \multicolumn{1}{c|}{\begin{tabular}[c]{@{}c@{}}0.61\\ (n=5)\end{tabular}}    & \begin{tabular}[c]{@{}c@{}}1.00\\ (n=12)\end{tabular} \\ \hline
50BVAGPT1               & \multicolumn{1}{c|}{0.97}                                                  & \multicolumn{1}{c|}{0.5}                                                   & \multicolumn{1}{c|}{0.97}                                                  & \multicolumn{1}{c|}{0.87}                                                  & \multicolumn{1}{c|}{0.94}                                                  & \multicolumn{1}{c|}{0.90}                                                     & \multicolumn{1}{c|}{0.64}                                                    & 1.00                                                  \\ \hline
50BVAGPT2               & \multicolumn{1}{c|}{0.91}                                                  & \multicolumn{1}{c|}{0.67}                                                  & \multicolumn{1}{c|}{0..97}                                                 & \multicolumn{1}{c|}{0.87}                                                  & \multicolumn{1}{c|}{0.94}                                                  & \multicolumn{1}{c|}{0.90}                                                     & \multicolumn{1}{c|}{0.24}                                                    & 1.00                                                  \\ \hline
50BVAGPT3               & \multicolumn{1}{c|}{0.88}                                                  & \multicolumn{1}{c|}{0.62}                                                  & \multicolumn{1}{c|}{1.00}                                                  & \multicolumn{1}{c|}{0.93}                                                  & \multicolumn{1}{c|}{0.94}                                                  & \multicolumn{1}{c|}{0.90}                                                     & \multicolumn{1}{c|}{0.66}                                                    & 1.00                                                  \\ \hline
50genGPT1                  & \multicolumn{1}{c|}{0.85}                                                  & \multicolumn{1}{c|}{0.68}                                                  & \multicolumn{1}{c|}{1.00}                                                  & \multicolumn{1}{c|}{0.93}                                                  & \multicolumn{1}{c|}{0.86}                                                  & \multicolumn{1}{c|}{0.90}                                                     & \multicolumn{1}{c|}{0.90}                                                    & 1.00                                                  \\ \hline
50genGPT2                  & \multicolumn{1}{c|}{0.85}                                                  & \multicolumn{1}{c|}{0.62}                                                  & \multicolumn{1}{c|}{1.00}                                                  & \multicolumn{1}{c|}{0.87}                                                  & \multicolumn{1}{c|}{0.86}                                                  & \multicolumn{1}{c|}{0.90}                                                     & \multicolumn{1}{c|}{0.90}                                                    & 1.00                                                  \\ \hline
50genGPT3                  & \multicolumn{1}{c|}{0.91}                                                  & \multicolumn{1}{c|}{0.64}                                                  & \multicolumn{1}{c|}{1.00}                                                  & \multicolumn{1}{c|}{0.93}                                                  & \multicolumn{1}{c|}{0.86}                                                  & \multicolumn{1}{c|}{0.90}                                                     & \multicolumn{1}{c|}{0.81}                                                    & 1.00                                                  \\ \hline
RT(n=50)                & \multicolumn{1}{c|}{0.76}                                                  & \multicolumn{1}{c|}{0.58}                                                  & \multicolumn{1}{c|}{0.79}                                                  & \multicolumn{1}{c|}{0.84}                                                  & \multicolumn{1}{c|}{0.47}                                                  & \multicolumn{1}{c|}{0.86}                                                     & \multicolumn{1}{c|}{0.87}                                                    & 0.95                                                  \\ \hline
MLBVA(n=50)             & \multicolumn{1}{c|}{0.88}                                                  & \multicolumn{1}{c|}{0.55}                                                  & \multicolumn{1}{c|}{0.91}                                                  & \multicolumn{1}{c|}{0.93}                                                  & \multicolumn{1}{c|}{0.77}                                                  & \multicolumn{1}{c|}{0.86}                                                     & \multicolumn{1}{c|}{0.16}                                                    & 0.95                                                  \\ \hline
Klee                    & \multicolumn{1}{c|}{\begin{tabular}[c]{@{}c@{}}1.00\\ (n=14)\end{tabular}} & \multicolumn{1}{c|}{\begin{tabular}[c]{@{}c@{}}0.74\\ (n=56)\end{tabular}} & \multicolumn{1}{c|}{\begin{tabular}[c]{@{}c@{}}1.00\\ (n=13)\end{tabular}} & \multicolumn{1}{c|}{\begin{tabular}[c]{@{}c@{}}0.28\\ (n=2)\end{tabular}}  & \multicolumn{1}{c|}{\begin{tabular}[c]{@{}c@{}}0.44\\ (n=4)\end{tabular}}  & \multicolumn{1}{c|}{\begin{tabular}[c]{@{}c@{}}1.00\\ (n=14221)\end{tabular}} & \multicolumn{1}{c|}{\begin{tabular}[c]{@{}c@{}}0.90\\ (n=1290)\end{tabular}} & \begin{tabular}[c]{@{}c@{}}1.00\\ (n=8)\end{tabular}  \\ \hline
\end{tabular}
}
\end{table*}

\begin{figure*}[htbp]
\centering
\includegraphics[width=0.95\linewidth]{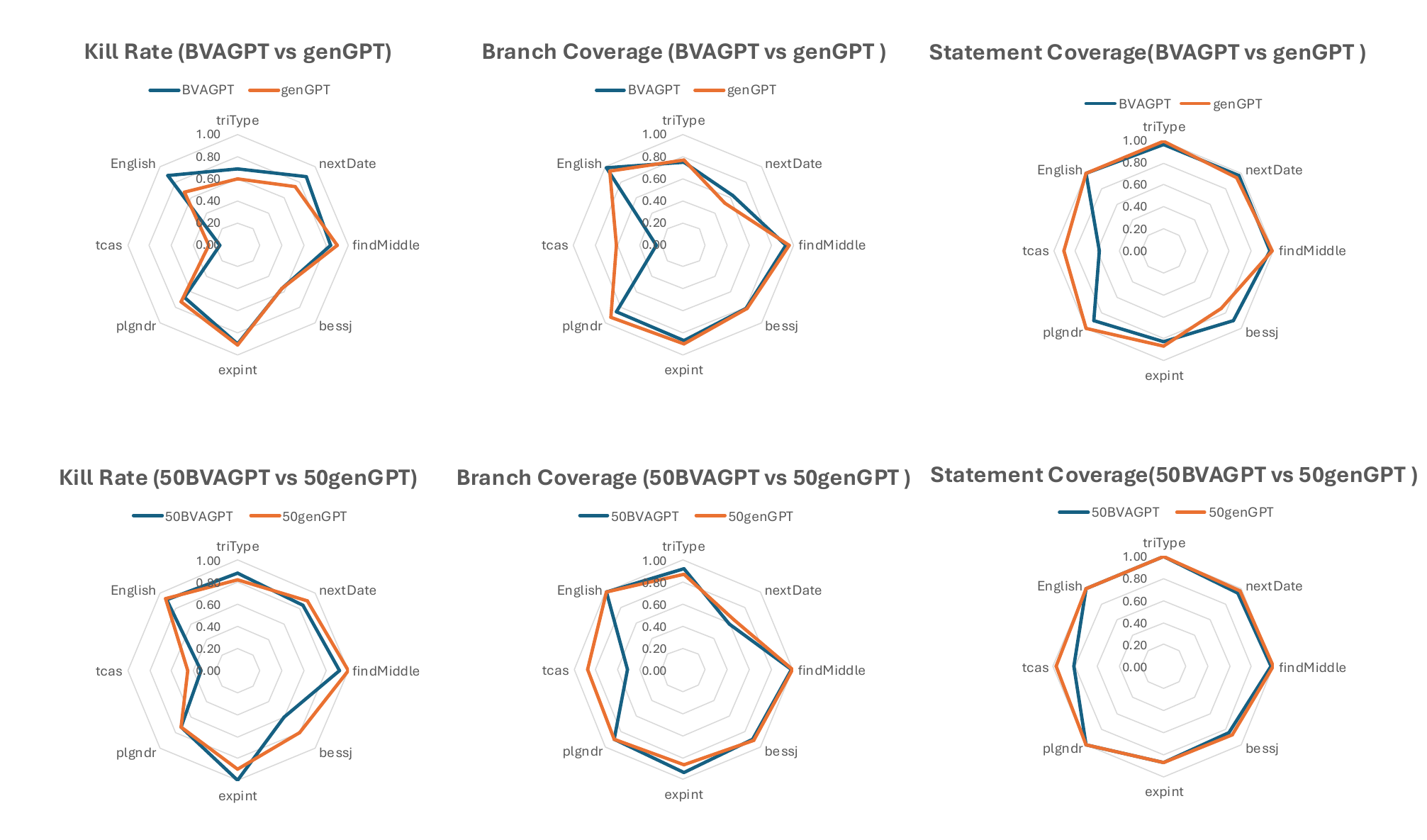}
\caption{Comparison of kill rates, statement and branch coverage for boundary and general test inputs across four prompts.}
\label{BVAvsGTP}
%\end{figure*}
%\begin{figure*}[htbp]
\hspace{100mm}
\centering

\includegraphics[width=1\linewidth]{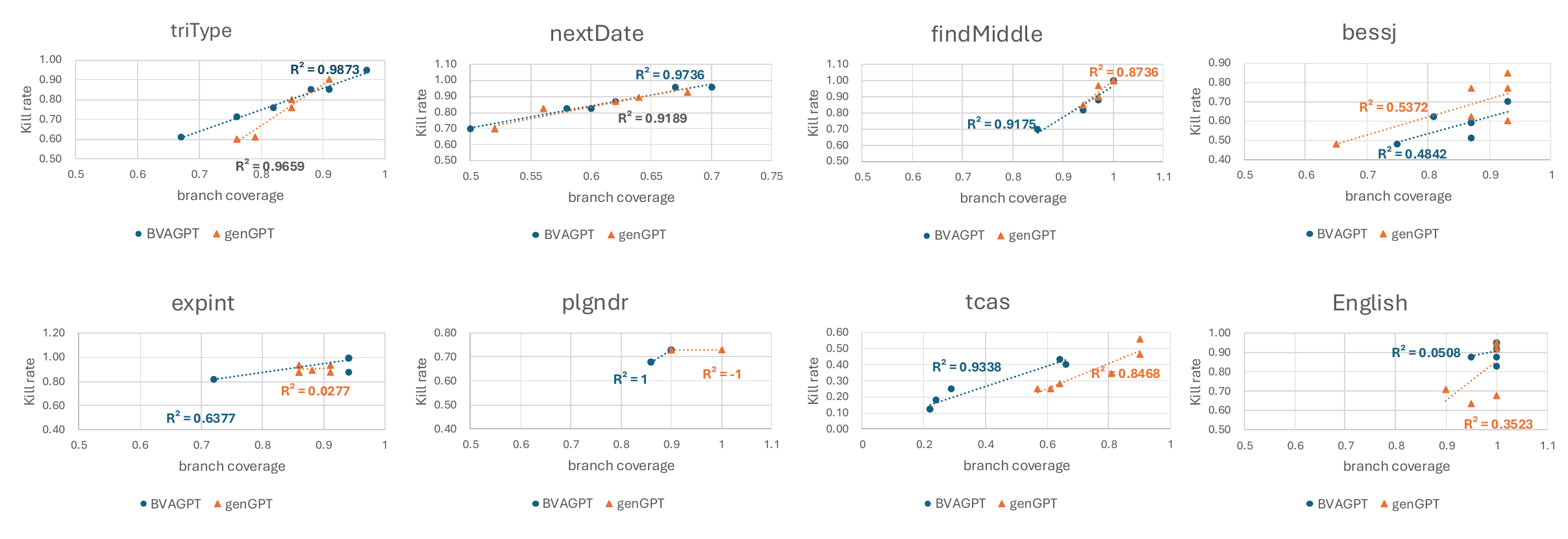}
\caption{Comparison of kill rates and branch coverage correlation between boundary-value and general test inputs.}
\label{covkill}
%\end{figure*}
%\begin{figure*}[htbp]
\centering

\hspace{150mm}
\includegraphics[width=1\linewidth]{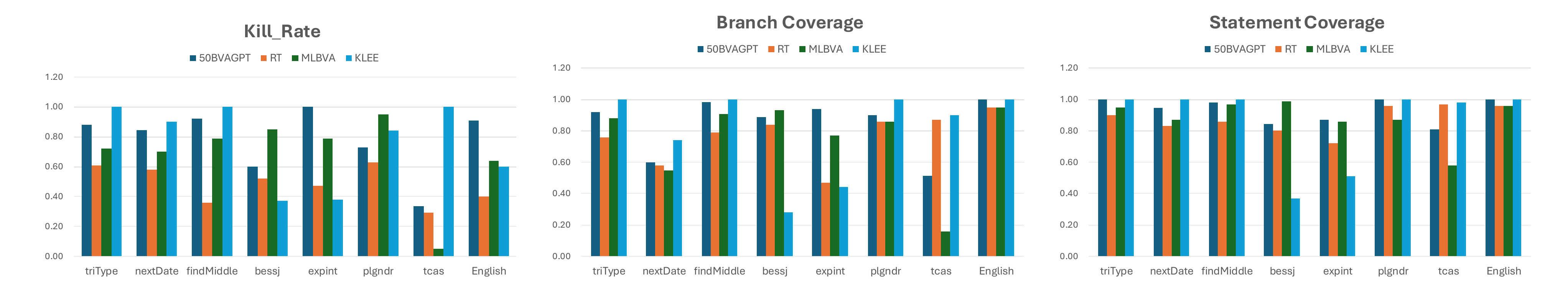}
\caption{kill rate, branch coverage, and statement coverage comparison analysis among 50BVAGPT, RT, MLBVA, and Klee.}
\label{GTPvsMLBVA}
%\end{figure*}
%\begin{figure*}[htbp]
%\centering
%\begin{minipage}{0.33\linewidth}
%\centering
%\includegraphics[width=1\linewidth]{figs/GTPvsMLBVA_kill.pdf}
%\caption{Kill rate comparison analysis among 50BVAGPT, RT, MLBVA, and Klee.}
%\label{GPTvsMLBVAkill}
%\end{minipage}%
%\begin{minipage}{0.33\linewidth}
%\centering
%\includegraphics[width=1\linewidth]{figs/GTPvsMLBVA_Scov.pdf}
%\caption{Statement coverage comparison among 50BVAGPT, RT, MLBVA, and Klee.}
%\label{GPTvsMLBVAScov}
%\end{minipage}%
%\begin{minipage}{0.33\linewidth}
%\centering
%\includegraphics[width=1\linewidth]{figs/GTPvsMLBVA_bcov.pdf}
%\caption{Branch coverage comparison among 50BVAGPT, RT, MLBVA, and Klee.}
%\label{GPTvsMLBVAbcov}
%\end{minipage}
\end{figure*}

%\begin{figure*}[tbp]
%\centering
%\includegraphics[width=1\linewidth]{figs/n_killrate.pdf}
%\caption{Kill rate and input count for boundary and general test inputs.}
%\label{nkillrate}
%\end{figure*}
%\begin{figure*}[tbp]
%\hspace{150mm}
%\centering
%\includegraphics[width=1\linewidth]{figs/n_segcov.pdf}
%\caption{Statement coverage and input count for boundary and general test inputs.}
%\label{nsegcov}
%\end{figure*}
%\begin{figure*}[tbp]
%\hspace{150mm}
%\centering
%\includegraphics[width=1\linewidth]{figs/n_branchcov.pdf}
%\caption{Branch coverage and input count for boundary and general test inputs.}
%\label{nbranchcov}
%\end{figure*}

\begin{figure*}[t]
\centering
\includegraphics[width=0.9\linewidth]{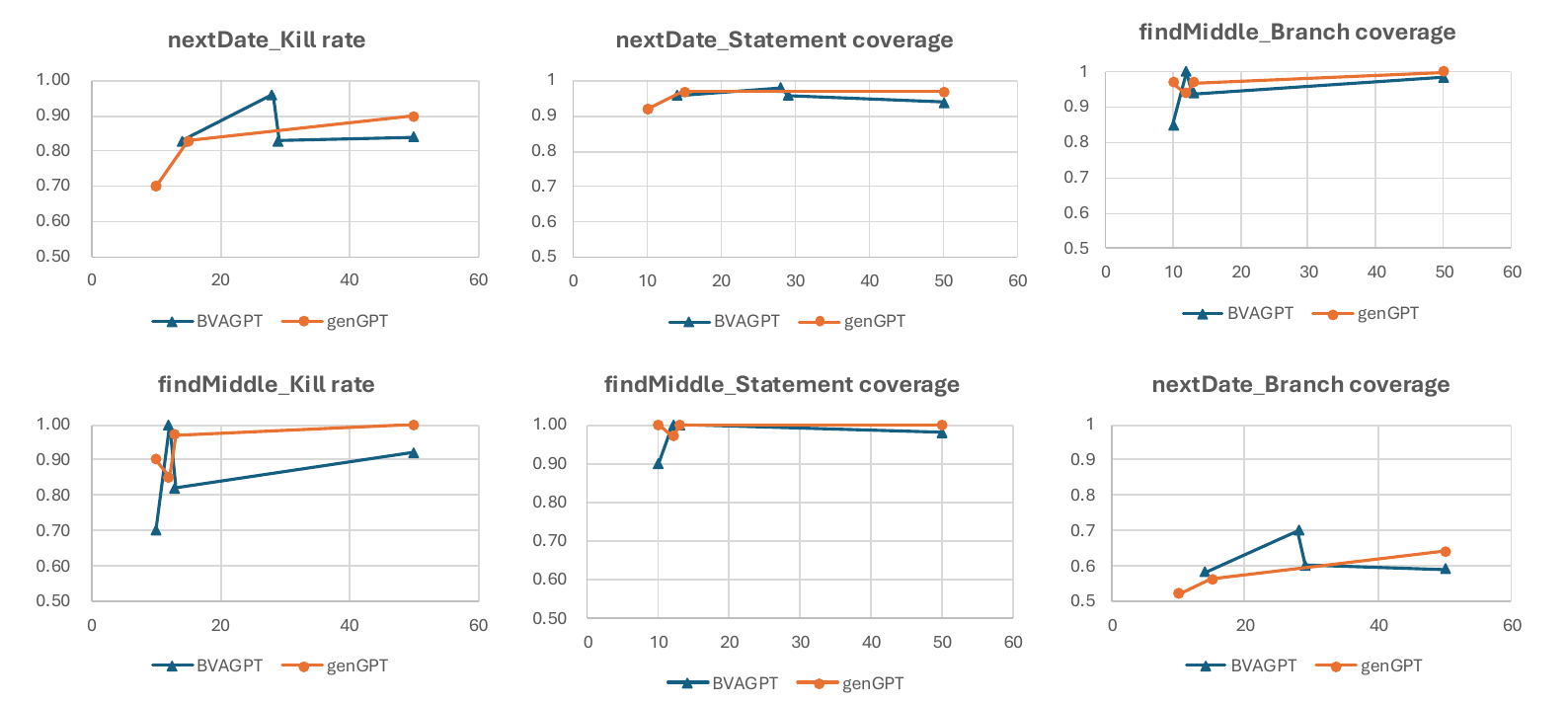}
\caption{Analysis of the relationship between input count, kill rate and coverage rate.}
\label{n_ksb}
\end{figure*}

\begin{figure}[t]
\centering
\includegraphics[width=0.85\linewidth]{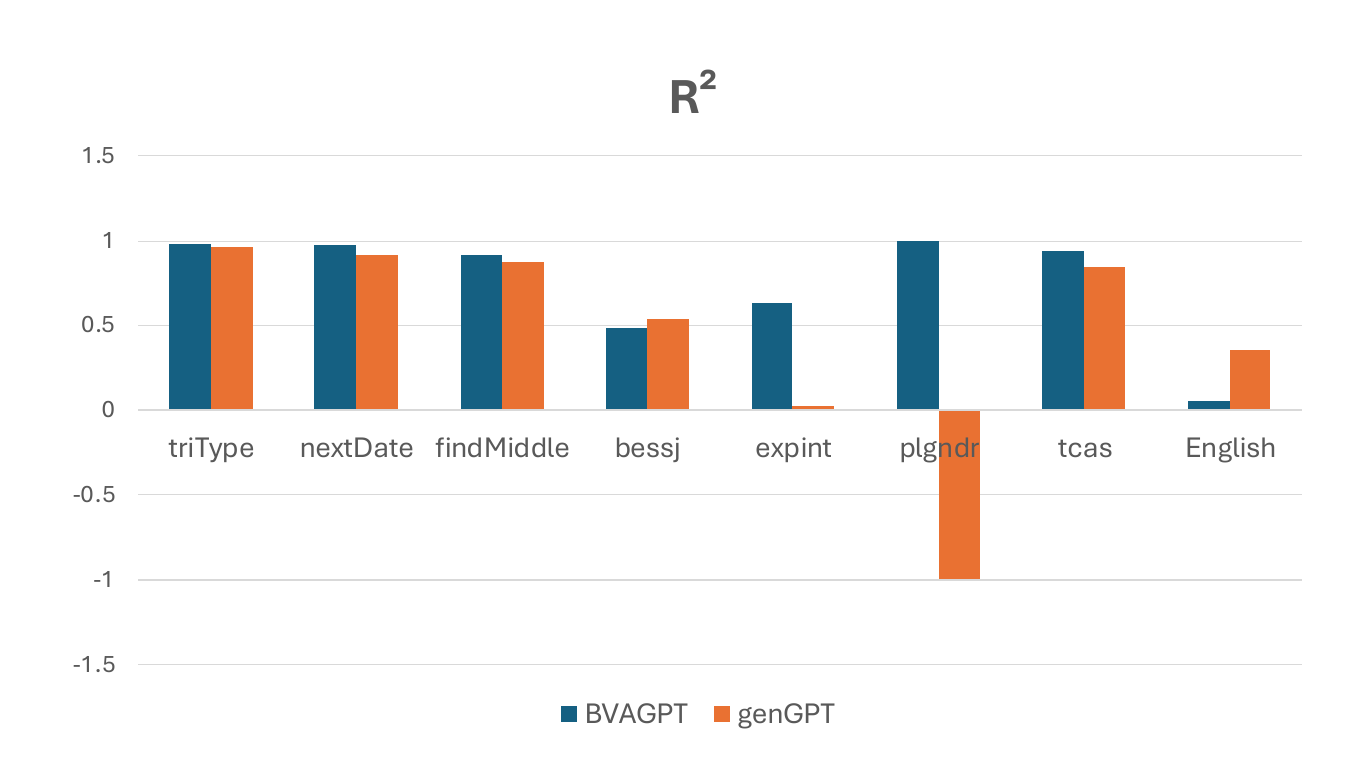}
\caption{$R^2$ correlation coefficient between branch coverage and fault detection rate across different programs.}
\label{R2}
\end{figure}

In this section, we conduct an experiment to evaluate the effectiveness of LLM-generated boundary value test inputs on fault detection and coverage rates for a set of eight test programs. The descriptions of these subject programs are shown in Table \ref{tab:table2}. And the details about all eight programs are shown in Table \ref{tab:table1}, such as the dimensional number of program inputs, the range of input domain, line of code (LOC), and injected fault information.

We selected GPT-4o as the LLMs for this study. We use four different prompts to guide GPT in generating test inputs, with each prompt designed to target either boundary value or general test inputs, allowing us to compare their effects on test effectiveness. Additionally, we compare the GPT-generated test inputs with those generated by three established testing methods: Random Testing (RT), concolic testing, and ML-based Boundary Value Analysis (MLBVA).

%~~~~~~~~~~~~~~~~~~~~~~~~~~~
\subsection{Baseline Test Input Generation Methods}
The \textit{RT} approach randomly generates a test set of 50 test inputs, which are then executed on each mutated program to assess fault detection effectiveness. 

\textit{Concolic testing}, also known as dynamic symbolic execution, is a hybrid verification technique that combines concrete execution with symbolic execution along a specified path. In symbolic execution, symbolic values replace standard inputs to a program, allowing the creation of a constraint set for each execution path. Afterward, constraint solvers analyze these constraints to determine which inputs trigger specific paths, aiming to maximize code coverage. Klee is a dynamic symbolic execution tool built on the LLVM framework, automates test input generation and achieves high program coverage \cite{cadar2008klee}. In this study, we apply Klee to generate test inputs for eight programs

\textit{MLBVA} \cite{Guo2023} is a machine learning-based approach for automatically generating boundary test inputs for software testing. The core idea is to train a machine learning classifier, or discriminator, to identify whether a boundary exists between two test inputs. Using the output from this discriminator, the method generates test inputs based on a Markov Chain Monte Carlo (MCMC) approach, effectively sampling inputs near the identified boundaries. In this experiment, we generated a training dataset of 10,000 samples for the discriminator, then used the trained model to generate 50 boundary-focused test inputs for each program under test.

%~~~~~~~~~~~~~~~~~~~~~~~~~~~
\subsection{Experimental Setup}
We designed four distinct prompts to instruct GPT in generating test inputs:

\begin{itemize}
\item {\bf Prompt 1:} Generate boundary value test inputs for c code delimited by triple backticks.
```
code
'''
 
\item {\bf Prompt 2:} Generate test inputs for c code delimited by triple backticks. 
```
code
'''
 
\item {\bf Prompt 3:} Generate 50 boundary value test inputs for c code delimited by triple backticks. 
```
code
'''
 
\item {\bf Prompt 4:} Generate 50 test inputs for c code delimited by triple backticks. 
```
code
'''
\end{itemize}

Prompts 1 and 3 are designed to guide GPT toward generating boundary value test inputs, while Prompts 2 and 4 focus on general test inputs, with the goal of comparing the effects of boundary-oriented and general test inputs on error detection. In Prompts 1 and 2, GPT autonomously determines the number of test inputs generated, while Prompts 3 and 4 are constrained to generate exactly 50 inputs.

Each prompt was executed three times per program, providing a diverse sample of inputs. Additionally, we generated 50 test inputs for each program using RT and MLBVA to compare these with the inputs from Prompt 3. We further used Klee to generate $n$ test inputs per program, with the exact values shown in Table \ref{tab:table3}.

For analysis purposes, we refer to the methods in our results as follows: BVAGPT for Prompt 1, genGPT for Prompt 2, 50BVAGPT for Prompt 3, and 50genGPT for Prompt 4. The experiment is designed to address the following key research questions (RQ):

\noindent{\bf RQ1:} Is there a significant difference in fault detection rates and coverage between boundary value and general test inputs generated by GPT?

\noindent{\bf RQ2:} Does the number of GPT-generated test inputs affect fault detection and coverage?

\noindent{\bf RQ3:} How do GPT-generated boundary value test inputs (Prompt  3) compare to traditional methods (RT, MLBVA, and concolic testing) in terms of fault detection and coverage?

\noindent{\bf RQ4:} Is there a positive correlation between fault detection rate and coverage?

%\noindent{\bf RQ5:} Does program complexity impact the effectiveness of GPT-generated test inputs?

\subsection{Result and Discussion}
\label{sec:discuss}
The results are summarized in three tables. Table \ref{tab:table3} compares the kill rate of test inputs generated by RT, Klee, MLBVA, and GPT, reflecting each method's effectiveness in detecting faults. Table \ref{tab:table4} shows the statement coverage achieved by each method, indicating the percentage of executed program statements. Table \ref{tab:table5} compares branch coverage, highlighting how well each method covers different program branches, particularly those involving boundary conditions. 

For {\bf RQ1}, we compare the average fault detection rates and coverage metrics (statement and branch coverage) across BVAGPT and genGPT, and similarly across 50BVAGPT and 50genGPT. This comparison aims to reveal any clear differences in test effectiveness between boundary-focused and general test inputs, particularly in the controlled 50-input case (50BVAGPT and 50genGPT).
Figure \ref{BVAvsGTP} shows the kill rates, statement coverage, and branch coverage for each of the four prompts, providing a visual comparison of boundary and general test input effectiveness. From the figure, it can be observed that when the number of test inputs is not specified, the kill rate for BVAGPT is higher than that of general test inputs for only three programs. However, when the number of inputs is increased, there is no significant difference in the kill rates between 50BVAGPT and 50genGPT. Upon examining the generated test inputs, we found that although GPT did not fully analyze boundary values when generating boundary-focused test inputs (prompt 1 and prompt3), it still considered some boundary values when generating general test inputs (prompt 2 and prompt 4), particularly when the number of inputs was set to 50. GPT prioritized coverage in generating general test inputs, explaining the lack of coverage advantage for boundary-focused prompts.

For {\bf RQ2}, we compare the number of test inputs generated across four prompts, and evaluate the relationship between test input quantity and corresponding kill rate and coverage rates. For example, Figure \ref{n_ksb} illustrates the results of the relationship analysis for the nextDate and findMiddle programs, assessing the impact of the number of test inputs on effectiveness. In BVAGPT (prompt 1) and genGPT (prompt 2), the results of each of the three executions per prompt are included in the figure, while for the 50-input cases (50BVAGPT and 50genGPT), the data represents the average of the three executions per prompt.

The figures show that test effectiveness does not consistently improve as the number of test inputs increases. This outcome suggests that simply increasing the quantity of inputs does not guarantee better fault detection or higher coverage. This result highlights a key challenge in using GPT to generate optimal test sets: finding the balance between quantity and quality. Generating an excessive number of test inputs can lead to diminishing returns, where the additional tests provide limited incremental benefit in coverage or fault detection. Conversely, a smaller, well-crafted set of inputs may capture essential scenarios more effectively. Future research should focus on refining prompt design and test generation strategies to prioritize impactful test inputs, aiming to achieve a more balanced and efficient test suite.

For {\bf RQ3}, Figure \ref{GTPvsMLBVA} shows the  mean comparisons to evaluate the kill rate, statement coverage, and branch coverage of test inputs generated by 50BVAGPT against those produced by RT, MLBVA, and Klee. The figures reveal that the 50 boundary-value inputs generated by GPT only surpassed the other methods in kill rate for two programs (expint and English). However, compared to the boundary-focused MLBVA method, GPT achieved higher kill rates in six out of the eight programs. For coverage metrics, the performance of GPT was mixed, with some programs showing strong results while others were outperformed by other methods. While GPT’s boundary-value generation does not consistently outperform other methods, it shows significant promise, particularly against MLBVA, in boundary-specific faults.

For {\bf RQ4}, Figure \ref{covkill} compares boundary-value test inputs with general test inputs and includes a regression analysis to assess the correlation between branch coverage and fault detection rates. For each generation method, we calculated the $R^2$ correlation coefficient to measure the relationship between branch coverage and fault detection rates. The experimental results indicate a positive correlation between fault detection rate and coverage: in six out of the eight programs, the boundary-value test inputs show a higher $R^2$ correlation coefficient than the general test inputs, as shown in Figure \ref{R2}. 

This positive correlation highlights the significance of boundary-value testing. While achieving 100\% branch coverage does not guarantee detection of boundary-specific faults, the higher fault detection rate observed with boundary-focused inputs suggests that these tests better target critical areas where errors commonly occur. This result underscores the unique value of boundary-value inputs, particularly in white-box testing scenarios, where covering boundary conditions in predicates can directly impact the effectiveness of fault detection. By focusing on boundary conditions, test inputs are better suited to uncover edge-case faults that might otherwise remain undetected with general test inputs, thus enhancing both fault detection and overall test reliability.

%Based on these results, future research in boundary value analysis (BVA) with LLMs should focus on enhancing boundary detection accuracy, optimizing the balance between test input quantity and quality, and refining prompt design to better capture specific boundary conditions. For instance, fine-tuning models with annotated boundary cases could improve effectiveness, and combining LLM-generated inputs with traditional methods like MLBVA and Klee may leverage the strengths of both approaches.
Future research in BVA with LLMs should focus on improving boundary detection accuracy, optimizing the balance between test input quantity and quality, and refining prompt design. Fine-tuning models with annotated boundary cases and combining LLM-generated inputs with traditional methods like MLBVA and Klee could enhance effectiveness.

%~~~~~~~~~~~~~~~~~~~~~~~~~~~~~~~~~~~~~~~~~~~~~~~~~~~~~~~~~
\section{Conclusion}
\label{sec:conclusion}

In this paper, we presented a framework for assessing the effectiveness of LLMs in generating boundary value test inputs for software testing, with a focus on white-box testing, by comparing LLM-generated inputs with traditional methods, including RT, MLBVA, and concolic testing. Our findings highlight that LLMs, when guided through specifically tailored prompts, show potential in generating boundary-focused test inputs that achieve comparable or even superior fault detection and coverage metrics to conventional techniques in certain cases. This study also explored how variations in prompt design-targeting either boundary or general inputs-impact the kill rate and coverage performance, providing insights into the interplay between input type and testing efficacy.

However, the results show that the increase in test input quantity does not guarantee improved effectiveness, indicating that the quality of test inputs remains a critical factor. Future work should focus on improving the accuracy of LLM-generated boundary values, refining prompt engineering to better capture boundary conditions, and exploring hybrid approaches that integrate LLM-based generation with traditional methods like MLBVA and KLEE. This study provides a foundational step towards leveraging LLMs for boundary value analysis in white-box testing, presenting both challenges and opportunities for enhancing software testing practices through AI-driven tools.

%~~~~~~~~~~~~~~~~~~~~~~~~~~~~~~~~~~~~~~~~~~~~~~~~~~~~~~~~~
%\section*{Acknowledgment}

%~~~~~~~~~~~~~~~~~~~~~~~~~~~~~~~~~~~~~~~~~~~~~~~~~~~~~~~~~
\bibliographystyle{IEEEtran}
%\newpage
\bibliography{refs}
\end{document}